\documentclass[pra,twocolumn,showpacs,superscriptaddress,floatfix, nofootinbib]{revtex4-1}
\usepackage[caption=false]{subfig}
\usepackage{amsmath,graphicx}
\usepackage{epsfig}
\usepackage{amsmath,amssymb,mathrsfs,esint}
\usepackage{gensymb}
\usepackage[bookmarks=true,
   colorlinks=true,
   linkcolor=blue,
   urlcolor=blue,
   citecolor=blue,
   bookmarks=true,
   hyperindex=true
]{hyperref}
\usepackage{soul}
\usepackage{natbib}
\usepackage{relsize}
\usepackage{float}
\usepackage{dsfont}

\usepackage{bm}

\newcommand{\be}{\begin{equation}}
\newcommand{\ee}{\end{equation}}

\newcommand{\beq}{\begin{eqnarray}}
\newcommand{\eeq}{\end{eqnarray}}

\def\H1{\widehat{H}_1}

\newcommand{\ket}[1]{\left| #1 \right>}

\begin{document}

\title{Free Fock parafermions in the tight-binding model with dissipation}

\author{A.\,S. Mastiukova}
\affiliation{Russian Quantum Center, Skolkovo, Moscow 143025, Russia}
\affiliation{Moscow Institute of Physics and Technology, Dolgoprudny, Moscow Region 141700, Russia}

\author{D.\,V. Kurlov}
\affiliation{Russian Quantum Center, Skolkovo, Moscow 143025, Russia}

\author{V. Gritsev}
\affiliation{Institute for Theoretical Physics Amsterdam, Universiteit van Amsterdam, Amsterdam 1098 XH, The Netherlands}
\affiliation{Russian Quantum Center, Skolkovo, Moscow 143025, Russia}

\author{A.\,K. Fedorov}
\affiliation{Russian Quantum Center, Skolkovo, Moscow 143025, Russia}
\affiliation{National University of Science and Technology ``MISIS”, 119049 Moscow, Russia}

\begin{abstract}
Parafermions that generalize (Majorana or usual) fermions appear as interacting quasi-particles because of their nature.  
Although attempts to develop models with {\it free} (non-interacting) parafermions have been undertaken, existing proposals require unphysical conditions such as realizing purely non-Hermitian systems. 
Here we present a way for the realization of free Fock parafermions in the tight-binding model with controlled dissipation of a very simple form. 
Introducing dissipation transforms an originally non-integrable quantum model to an exactly solvable classical one.

\end{abstract}

\maketitle

\section{Introduction} 
The combination of fundamental importance and potential applications puts analyzing quantum systems hosting exotic quasiparticle excitations to the forefront of ongoing research. 
Specifically, anyons that exhibit richer exchange statistics have received a significant attention~\cite{Bogoliubov1992,Eckern1998,Eckern20001,Eckern20002,Kundu1999,Oelkers2006,Mintchev2007,Cabra2007,Pucci2007,Averin2007,Patu2015,Freedman2007,Troyer2008,Wang2009,Calabrese2008,Min2008,Mintchev2009,Roncaglia2011,Korepin2012,Santos2015,Pelster2015,Hao2016,Calabrese2016,Valencia2016,Valencia2018,Trombettoni2018}. 
Potential applications of non-Abelian anyons are mainly related to the usage of their braiding statistics for topologically protected quantum information processing~\cite{Nayak2008}. 
Finding an experimentally relevant setup to host such exotic quasiparticles is a challenge. 
One of the options is to use systems, in which non-Abelian anyons arise from Majorana zero modes bound to vortices~\cite{Meyer2011,Oppen2011,Furdyna2012,Yacoby2014,Aguado2012}, 
such as a fractional quantum Hall state at $5/2$ filling~\cite{MooreRead1991}, 
unconventional $p+ip$ superconductors~\cite{Volovik1999,Green2000,Liu2004,Rice2004,Tewari2006}, and semiconductor wires~\cite{DasSarma2010,Sau2010,Oppen2010,Alicea2010,Fisher2011}.
However, it is impossible to realize the universal set of quantum gates in such cases since they cannot be approximated by braiding operations~\cite{Fisher2011,Oppen2012,LossKlinovaja2021}.
Several proposals to overcome this difficulty have been suggested~\cite{Freedman2006,Bonderson2010}, but realistic ones require non-topological operations killing the main feature of such systems --- their immunity from decoherence~\cite{Nayak2008}.

An important step is to go beyond existing models to the ones having more complex braiding statistics and allowing to manipulate quantum states more efficiently. 
We note that in addition to solid-state systems, programmable quantum simulators can be used for studying exotic quasiparticles. 
In particular, Rydberg quantum simulator allows realizing Ising-type models~\cite{Lukin2017,Lukin2021} and more complex ${\mathbb Z}_n$ lattice models related to the $n$-state Potts or ${\mathbb Z}_n$ chiral clock model.
Recently, Rydberg ensembles have been studied in the context of mesonic and baryonic quasiparticle excitations in the ${\mathbb Z}_3$ case~\cite{Gorshkov2020}, and more intriguing features could appear in the ${\mathbb Z}_n$ case. 
We note that lattice models based on a discrete group ${\mathbb Z}_{n}$ were discovered in Ref.~\cite{Andrews1984}. Critical properties of these models are described by a family of minimal models of 2D conformal field theories which possesses current fields with non-integer spins, called parafermions in Ref.~\cite{Zamolodchikov1985}.


Over the past decade parafermions have been a subject of intensive research~\cite{Fendley2012,Fendley2014,Fendley2016,KlinovajaLoss2014} 
in particular from the perspective of quantum computing~\cite{HutterLoss2016}. 
Specifically, there is a significant interest to Fock parafermions (FPF), which are anyonic quasi-particles that generalize usual {\it identical} fermions in a simple particle-like picture~\cite{Cobanera2014}
(standard parafermions generalize Majorana fermions). 
Various properties of standard and Fock parafermions~\cite{Hassler2016,Mazza2017,Alicea2018,Cobanera2017} 
and schemes for their realization~\cite{Fendley2016,Lindner2012,Burrello2013,Mong2013,Barkeshli2014,ZhangKane2014,Clarke2013,LossKlinovaja2021,KlinovajaLoss20142,Schmidt2015,KlinovajaLoss20143,KlinovajaLoss20144,Loss2015,KlinovajaLoss2015,Thakurathi2017,Laubscher2019,Laubscher2020,Ronetti2021} 
have been considered.
The simplest setup for Fock parafermions is a one-dimensional (1D) tight-binding model, 
which is however shown to be interacting and non-integrable~\cite{Rossini2019}. 
Numerical simulations based on exact diagonalization and on the density-matrix renormalization group have predicted bound states in the spectrum due to its peculiar anyonic properties~\cite{Rossini2019}. 
Although considerable efforts have been made in searching for free (i.e. non-interacting) parafermionic models~\cite{Fendley2014}, existing proposal require implementing purely non-Hermtian systems.
For instance, this is the case with the Baxter non-Hermitian Hamiltonian~\cite{Baxter1989}, which possesses a free-particle complex spectrum. 
Many properties of the Baxter Hamiltonian have been discussed in Ref.~\cite{Fendley2014}.
It should be noted that in the case of standard parafermions, free particle behaviour may take place in the the low-energy sector of some models~\cite{Zamolodchikov1985,Ronetti2021}.
However, the question of the very existence of physically relevant models hosting free FPF remains open.

Here we positively answer to this question by demonstrating a 1D model with free FPF with dissipation. 
Including dissipation of a very simple form transform a non-integrable model of Fock parafermion to an exactly-solvable one by natural canceling the Hermitian conjugate part of the Baxter Hamiltonian. 
We expect that experimental realization of the suggested model should be possible using solid-state realization and programmable quantum simulators.

\section{Fock parafermions} 
Let us make a brief overview of Fock parafermoins and some of their properties before discussing the model. FPFs are anyonic quasi-particles that generalize usual identical fermions. First of all, the difference is in the dimensionality of the Fock space. Whereas for fermions the local Hilbert space (say, on a lattice site) is two dimensional, for~${\mathbb Z}_n$ FPFs it is~$n$-dimensional. Thus, introducing the FPF creation and annihilation operators,~$F_j^{\dag}$ and~$F_j$ correspondingly, one has 
\be \label{FPF_number_state}
	(F_j^{\dag})^m \ket{0} = \ket{m_j}, \qquad 0\leq m \leq n-1,
\ee
where $\ket{0}$ is a vacuum, $\ket{m_j}$ is a Fock state with~$m$ FPFs on the~$j$-th site and nothing on the rest of the chain. In other words, one can have up to~$n-1$ identical FPFs in the same state. For a general Fock state, the action of $F_j^{\dag}$ and $F_j$ is a bit more involved~\cite{Cobanera2014}:
\be \label{many_body_Fock_state_FPF_ops_action}
\begin{aligned}
	&F_j^{\dag} \ket{ \ldots, m_j, \ldots } = \omega^{- \sum_{k = 1}^{j-1} m_k } \ket{ \ldots, m_j + 1, \ldots },\\
	&F_j \ket{ \ldots, m_j, \ldots } = \omega^{\sum_{k = 1}^{j-1} m_k } \ket{ \ldots, m_j - 1, \ldots },
\end{aligned}
\ee
where~$\omega = e^{2\pi i /n}$.
FPF creation and annihilation operators satisfy the following relations:
\be \label{FPF_rels1}
	(F_j^{\dag})^n = F_j^n = 0, \quad (F_j^{\dag})^m F_j^m + F_j^{n-m} (F_j^{\dag})^{n-m} = 1,
\ee
with~$1 \leq m \leq n-1$. As an immediate consequence of Eq.~(\ref{FPF_rels1}), one has $F_j^{m} (F_j^{\dag})^m F_j^m = F_j^m$. For operators acting on different sites, one has
\be \label{FPF_braiding_rels} 
	F_j F_k = \omega^{\text{sgn}(k-j) } F_k F_j, \quad F_j^{\dag} F_k = \omega^{- \text{sgn}(k-j) } F_k F_j^{\dag},
\ee
where $\text{sgn}(x)$ is the sign function. Thus, Eqs.~(\ref{FPF_rels1}) and~(\ref{FPF_braiding_rels}) allow one to bring any monomial in FPFs to the normal order with all creation operators being to the left from all annihilation operators. Further, one introduces the FPF number operator
\be \label{number_op}
	N_j = \sum_{m=1}^{n-1} (F_j^{\dag})^m F_j^m,
\ee
which acts on the Fock states as expected:
\be
	N_j \ket{ \ldots, m_j, \ldots} = m_j \ket{ \ldots, m_j, \ldots },
\ee
and satisfies the usual commutation relations
\be \label{N_F_comm}
	\left[ N_j, F_j \right] = - F_j, \qquad	[ N_j, F_j^{\dag} ] = F_j^{\dag}.
\ee
One can easily check that Eqs.~(\ref{FPF_rels1})--(\ref{number_op}) reduce to the standard fermionic relations for~$n=2$. 
On the other hand, for~$n>2$, from~Eq. (\ref{FPF_braiding_rels}) one clearly sees that FPFs are anyons with the statistical parameter~$2/n$. In this case, the factor of~$\omega^{\text{sgn}(k-j)}$ in Eq.~(\ref{FPF_braiding_rels}) comes into play, which makes a crucial difference between FPFs and fermions and has far-reaching consequences. For instance, it makes FPFs intrinsically interacting and the Fourier components of~$F_j$, defined as 
\be
	F_k = \frac{1}{\sqrt{L}} \sum_{j = 1}^L e^{i k j} F_j,
\ee
are not FPFs themselves, as they {\it do not} satisfy the relations~(\ref{FPF_braiding_rels}). As a result, even if some quadratic FPF Hamiltonian can be written in the momentum space as~$H = \sum_k \varepsilon_k F_k^{\dag} F_k$, it is not diagonal since different momentum modes are coupled. Indeed, it is easy to check that the commutator $[F_k^{\dag} F_k, F_q^{\dag} F_q]$~ is not zero~\cite{Rossini2019}.

Just like the usual spinless fermions are related to the spin-$1/2$ Pauli operators via the Jordan-Wigner transformation, FPFs can be obtained from the generalized~${\mathbb Z}_n$ Pauli operators~$X_j$ and~$Z_j$.
The latter are unitary and satisfy
\be
	X_j^n = Z_j^n  = 1, \quad X_j^{\dag} = X_j^{n-1}, \quad Z_j^{\dag} = Z_j^{n-1},
\ee 
along with the following relations:
\be \label{X_Z_properties}
	X_j^k Z_j^l = \omega^{k l} Z_j^l X_j^k, \quad X_j Z_k = Z_ k X_j \;\; (j\neq k).
\ee
The operators~$X_j$ and~$Z_j$ act non-trivially on $j$-th site and their matrix representations are $X_j = 1\otimes \ldots 1 \otimes X \otimes1 \ldots$ and $Z_j = 1\otimes \ldots \otimes Z \otimes \ldots$, where $X$ and $Z$ are~$n \times n$ matrices with the following elements:  
$Z_{p,q} = \delta_{p,q} \omega^{p-1}$ and~$X_{p,q} = \delta_{(p \text{ mod } n) + 1, q}$.
Then, using a generalization of the Jordan-Wigner transformation, the so-called Fradkin-Kadanoff transformation~\cite{Fradkin1980}, for the FPF annihilation operator one obtains
\be \label{F_to_Z_Sigma}
	F_j = \left( \prod_{k=1}^{j-1} Z_k \right) \Sigma_j^{-},
\ee
where~$\Sigma_j^{-}$ is the lowering operator. In terms of~$X_j$ and~$Z_j$ the latter is given by
\be \label{Sigma_minus}
	\Sigma_j^{-} = \frac{n-1}{n} X_j   -  \frac{1}{n}X_j \sum_{m=1}^{n-1} Z_j^m.
\ee
Its matrix representation is~$\Sigma_j^- = 1\otimes \ldots 1\otimes \Sigma^{-} \otimes1 \ldots $, with the elements of~$\Sigma^-$ being  $\Sigma^{-}_{p.q} = \delta_{p+1, q}$.

For completeness, we briefly comment on the notion of {\it standard} parafermions.  
We stress once more that {\it Fock} parafermions generalize usual complex fermions, which are related to real Majorana fermions by a linear transformation. In turn, Majorana fermions are generalized by  parafermions~$\gamma_j$. The latter satisfy the generalized Clifford algebra relations
\be \label{gamma_parafermions_propertires}
	\gamma_{j}^n = 1, \quad \gamma_j^{\dag} = \gamma_j^{n-1}, \quad \gamma_j \gamma_k = \omega^{\text{sgn}(k-j)} \gamma_k \gamma_j,
\ee
 and can be written in terms of~${\mathbb Z}_n$ matrices as
 \be \label{gamma_to_Z}
 	\gamma_{2j-1} = \left( \prod_{k=1}^{j-1} Z_k \right) X_j, \quad \gamma_{2j} = \omega^{(n-1)/2} \gamma_{2j-1} Z_j.
 \ee
 The relation between Fock parafermions~$F_j$ and parafermions~$\gamma_j$ can be easily obtained from Eqs.~(\ref{F_to_Z_Sigma})--(\ref{gamma_to_Z}), and it reads as
\be
	F_j = \frac{n-1}{n} \gamma_{2j-1} - \frac{1}{n} \sum_{m=1}^{n-1} \omega^{m^2 / 2 } \left(\gamma_{2j-1}^{m-1}\right)^{\dag} \gamma_{2j}^m,
\ee
were we took into account that $Z_j^m = \omega^{m^2/2} \gamma_{2j-1}^{\dag \, m} \gamma_{2j}$, as follows from~Eq.~(\ref{gamma_to_Z}).
Let us also mention that one can construct FPF coherent states (eigenstates of the annihilation operator~$F_j$) using an appropriate generalization of Grassmann variables, see e.g. Ref.~\cite{Trifonov2012}.

\section{The Model} 
We study the FPF tight-binding model on a chain of~$L$ sites with open boundary conditions. Since one can have up to~$n-1$ FPFs on each lattice site, we allow hopping processes involving~$1 \leq m \leq n-1$ particles, generalizing a usual single-particle hopping.
The Hamiltonian reads
\be \label{H}
\begin{aligned}
	H =& - \sum_{m=1}^{n-1} \sum_{j = 1}^{L-1} \left( \alpha_m (F_{j}^{\dag})^m F_{j+1}^m + \text{H.c.}  \right) \\
	& - \sum_{m=1}^{n-1} \mu_m \sum_{j=1}^L (F_j^{\dag})^m F_j^m,
\end{aligned}
\ee
where~$\alpha_m$ is the~$m$-particle hopping amplitude and $\mu_m$ is the chemical potential which depends on the number of FPFs on the site. 
For the moment we allow~$\alpha_m$ to be complex.
Naively, the model in~Eq.~(\ref{H}) looks quite simple. However, it turns out to be fairly complicated due to the non-trivial relations~(\ref{FPF_braiding_rels}) satisfied by~$F_j$ and~$F_k^{\dag}$. Indeed, consider first the simplest case with~$\mu_m = 0$ and only a single-particle hopping ($\alpha_m= 0$ for $m > 1$), such that the Hamiltonian~(\ref{H}) is quadratic. However, it does {\it not} describe free particles. Moreover, it is {\it not integrable}, as shown in Ref.~\cite{Rossini2019} for both open and periodic boundary conditions. The same holds in the general case: the Hamiltonian~(\ref{H}) with arbitrary values of~$n$, $\alpha_m$, and~$\mu_m$ is neither free nor integrable. Nevertheless, it exhibits rich physics, and for $n = 3$ and $\mu_m = 0$ the phase diagram was obtained in Ref.~\cite{Mahyaeh2020}. 
Quite remarkably, the model~(\ref{H}) simplifies dramatically in the presence of dissipation, as we shortly demonstrate. 

We now assume that the system~(\ref{H}) is not perfectly isolated, but is coupled to a Markovian dissipative environment. In this case one has to deal with the Lindblad master equation, which we write as (in what follows we set $\hbar = 1$)
\be \label{Lindblad_eq}
	\partial_t \rho = -i \left( H_{\text{eff}} \, \rho - \rho\, H_{\text{eff}}^{\dag} \right) + \sum_{j, m} \gamma_m \, L_{j}^{(m)} \, \rho\, L_{j}^{(m) \, \dag} \equiv {\cal L} \rho,
\ee 
where $\rho$ is the density matrix, ${\cal L}$ is the Liouvillian superoperator, $L_{j,q}$ is the jump operator acting on the $j$-th site, $\gamma_q > 0$ is a constant representing the dissipation strength, and the index~$q$ labels different types of dissipative processes. In Eq.~(\ref{Lindblad_eq}) we introduced an effective non-Hermitian Hamiltonian
\be \label{H_eff_gen}
	H_{\text{eff}} = H - i \,  \sum_{j , m} \frac{\gamma_m}{2}  L_{j}^{(m)\, \dag} L_{j}^{(m)},
\ee 
with the Hermitian part~$H$ given by~Eq.~(\ref{H}).  We then take the jump operators of the following form:
\be \label{jump_op}
	L_{j}^{(m)} = F_j^m + \Delta_m F_{j+1}^m, \qquad 1 \leq m \leq n-1,
\ee
where~$\Delta_m \neq 0$ is a complex parameter. The jump operator~(\ref{jump_op}) is (quasi)local and acts on a single link between two adjacent sites~\cite{Ziolkowska2020}. We choose the values of~$\Delta_m$ that satisfy
\be
	\alpha_m = i \gamma_m \Delta_m /2.
\ee
Then,~Eqs.~(\ref{H}) and ~(\ref{H_eff_gen}) yield
\be \label{H_eff}
\begin{aligned}
	 &H_{\text{eff}} = - \sum_{m=1}^{n-1} \Bigl\{ i \gamma_m \Delta_m \sum_{j=1}^{L-1} (F_j^{\dag})^m F_{j+1}^m  \\
	&+ i \, \frac{\gamma_m}{2} \sum_{j=1}^{L-1} \left( (F_j^{\dag})^m F_j^m + |\Delta_m|^2 (F_{j+1}^{\dag})^m F_{j+1}^m \right) \\
	&+ \mu_m \sum_{j=1}^L (F_j^{\dag})^m F_j^m \Bigr\}. 
\end{aligned}
\ee
Note that the effective Hamiltonian~(\ref{H_eff}) has a very simple structure. Indeed, taking into account~Eq.~(\ref{F_to_Z_Sigma}) we  have~$(F_{j}^{\dag})^m F_{j+1}^m = (\Sigma_j^{+})^m Z_j^m (\Sigma_{j+1}^{-})^m$ and $(F_j^{\dag})^m F_j^m = (\Sigma_j^{+})^m (\Sigma_j^{-})^m $, with~$\Sigma_j^{+}= (\Sigma_j^{-})^{\dag}$ being the raising operator. Then, using the expressions for the matrix elements of~$Z_j$ and~$\Sigma_j^{-}$ written after~Eqs.~(\ref{X_Z_properties}) and~(\ref{Sigma_minus}) correspondingly, we see that~$H_{\text{eff}}$ has a lower-triangular structure in the Fock basis. Moreover, the term on the first line of~Eq.~(\ref{H_eff}) does not have diagonal matrix elements, whereas the remaining terms [the second and the third lines in Eq.~(\ref{H_eff})] are diagonal and contribute to the eigenvalues. Therefore, the spectrum of~$H_{\text{eff}}$ can be readily obtained and it reads
\be
	H_{\text{eff}}\ket{m_1, \ldots, m_L} = E_{ m_1, \ldots, m_L } \ket{m_1, \ldots, m_L},
\ee
where~$\ket{m_1, \ldots, m_L}$ with~$0\leq m_j \leq n-1$ form the basis in the $n^L$-dimensional Fock space and the eigenvalues are given by
\be \label{H_eff_spectrum}
	E_{ m_1, \ldots, m_L }  \equiv E_{\boldsymbol{m}} = - \sum_{j=1}^{L} \sum_{p = 1}^{m_j} \left( \mu_p + \frac{ i }{2} \Gamma_{j}^{(p)} \right),
\ee
where~$\Gamma_1^{(p)} = \gamma_p$, $\Gamma_L^{(p)} = \gamma_p |\Delta_p|^2$, and $\Gamma_j^{(p)} = \Gamma_1^{(p)} + \Gamma_L^{(p)}$ for $2 \leq j \leq L-1$.  
As expected, the spectrum~(\ref{H_eff_spectrum}) is complex, except for the ground state, since~$H_{\text{eff}}$ is neither Hermitian nor ${\cal PT}$-symmetric~\cite{PTsymmetry}.
Remarkably, we see that Eq.~(\ref{H_eff_spectrum}) is nothing else than the spectrum of free Fock parafermions, since it is a linear combination of single-particle energy levels.  
Moreover, as we show below, the spectrum and the eigenstates of the full Liouvillian possess the same free-particle structure.

\section{Liouvilian spectrum and free Fock Parafermions}
The non-Hermitian Hamiltonian~$H_{\text{eff}}$ from~Eq.~(\ref{H_eff_gen}) is used routinely in the so-called quantum trajectories technique, which numerically solves a stochastic differential equation for a wavefunction, rather than for a density matrix~\cite{Dalibard1992, Gardiner1992, Plenio1998,Daley2014}. The quantum trajectories technique is quite successful in obtaining steady state averages of local observables. However, this method provides a rather poor approximation of an open system described by the full Liouvillian~${\cal L}$, if one is interested in e.g. the spectral properties. Interestingly, there are some exceptional cases in which~$H_{\text{eff}}$ does provide a complete description of a dissipative quantum system. For instance, this situation occurs if the following conditions are met: ({\it i\,}) there is an observable~$Q$ that commutes with the Hamiltonian, $[ H, Q] = 0$; ({\it ii\,}) all jump operators~$L_j$ correspond to either pure loss or pure gain, and ({\it iii\,}) jump operators satisfy~$[ L_j , Q] \propto L_j$. In this case the the Liouvillian spectrum is given by $\lambda_{m,n} = - i (E_n - E_m^*)$,
where~$E_n$ belong to the energy spectrum of the non-Hermitian Hamiltonian~(\ref{H_eff_gen}). In the same way the Liouvillian eigenstates are constructed from the eigenstates of~$H_{\text{eff}}$~\cite{Torres2014}.
Recently, an exact solution for the dissipative one-dimensional Hubbard model was obtained with the help of the outlined approach~\cite{Nakagawa2021}. 
Importantly, it is also applicable in our case. Indeed, the Hamiltonian~(\ref{H}) possesses the $U(1)$ symmetry and conserves the number of particles since~$[ H, {\cal N} ] = 0$, where ${\cal N} = \sum_{j=1}^L N_j$ is the total number operator. Further, 
the jump operator~$L_j$ from Eq.~(\ref{jump_op}) clearly describes pure losses. Finally, taking into account Eq.~(\ref{N_F_comm}), we see that~$[{\cal N}, L_j] = - L_j$. Therefore, for the Liouvillian~${\cal L}$ from Eq.~(\ref{Lindblad_eq}), with the Hamiltonian~(\ref{H}) and the jump operator~(\ref{jump_op}), we immediately find the spectrum that reads
\be \label{Liouvillian_spectrum}
	\lambda_{ {\boldsymbol m}, {\boldsymbol p} } = - i \left( E_{\boldsymbol{m}} - E_{\boldsymbol{p}}^* \right),
\ee
where~$E_{\boldsymbol{m}}=E_{m_1,\ldots m_L}$ is given by Eq.~(\ref{H_eff_spectrum}). Similarly, one can obtain the Liouvillian eigenstates from the eigenstates of~$H_{\text{eff}}$~\cite{Torres2014}. 

Let us discuss the spectrum~(\ref{Liouvillian_spectrum}) in more detail. First of all, taking into account Eq.~(\ref{H_eff_spectrum}) we see that there is always a unique zero eigenvalue~$\lambda_{\boldsymbol{0}, \boldsymbol{0} } = 0$. For $\gamma_1 \neq 0$, the remaining Liouvillian eigenvalues have a {\it finite} negative real part, which means that the steady state of Eq.~(\ref{Lindblad_eq}) is a vacuum. 
However, the situation drastically changes if in Eq.~(\ref{H}) we take~$\alpha_1 = 0$ and in Eq.~(\ref{H_eff}) put~$\gamma_1 = 0$. In other words, we forbid the single-particle hopping and losses. In this case it follows from Eqs.~(\ref{H_eff_spectrum}) and~(\ref{Liouvillian_spectrum}) that the states with no more than one particle per lattice site form a decoherence-free subspace consisting of~$2^L - 1$ ``dark states''. 
 
\section{Conclusions}
To summarize, we have found a simple open quantum system that hosts free (non-interacting) Fock parafermions. Starting from the non-integrable FPF tight-binding model~(\ref{H}) and including a Markovian dissipation described by the jump operator of a very simple form~(\ref{jump_op}), we have shown that for a specific values of the system parameters the Liouvillian spectrum has a free-particle structure and is given by Eqs.~(\ref{H_eff_spectrum}) and (\ref{Liouvillian_spectrum}). In contrast to all previously known free ${\mathbb Z}_n$ models, we had accurately and consistently treated the dissipation by working with the Lindblad equation~(\ref{Lindblad_eq}). Thus, our results provide a realistic physical system that hosts exotic and long-sought free Fock parafermions.
We expect that our predictions can be probed on the basis of currently available experimental
facilities using solid-state systems or programmable quantum simulators. 

\acknowledgements
We thank D. Loss and M. Lukin for useful discussions and valuable contribution at the early stages of the work during ICQT-2021.
This work is supported by the Russian Science Foundation No. 20-42-05002 and Russian Roadmap for Quantum Computing.
The work by VG is part of the DeltaITP consortium, a
program of the Netherlands Organization for Scientific Research (NWO) funded by the Dutch Ministry
of Education, Culture and Science (OCW).


\begin{thebibliography}{99}

\bibitem{Bogoliubov1992}
N. M. Bogoliubov and R. K. Bullough,
A q-deformed completely integrable Bose gas model,
\href{https://doi.org/10.1088/0305-4470/25/14/020}{J. Phys. A: Math. Gen. {\bf 25}, 4057 (1992)}.

\bibitem{Eckern1998}
L. Amico, A. Osterloh, and U. Eckern,
One-dimensional XXZ model for particles obeying fractional statistics,
\href{https://doi.org/10.1103/PhysRevB.58.R1703}{Phys. Rev. B {\bf 58}, R1703(R) (1998)}.

\bibitem{Eckern20001}
A. Osterloh, L. Amico, and U. Eckern,
Fermionic long-range correlations realized by particles obeying deformed statistics,
\href{https://doi.org/10.1088/0305-4470/33/48/104}{J. Phys. A: Math. Gen. {\bf 33}, L487 (2000)}.

\bibitem{Eckern20002}
A. Osterloh, L. Amico, and U. Eckern,
Exact solution of generalized Schulz-Shastry type models,
\href{https://doi.org/10.1016/S0550-3213(00)00496-X}{Nucl. Phys. B {\bf 588}, 531 (2000)}.

\bibitem{Kundu1999}
A. Kundu,
Exact solution of double $\delta$ function Bose gas through an interacting anyon gas,
\href{https://doi.org/10.1103/PhysRevLett.83.1275}{Phys. Rev. Lett. {\bf 83}, 1275 (1999)}.

\bibitem{Oelkers2006}
M. T. Batchelor, X.-W. Guan, and N. Oelkers,
One-dimensional interacting anyon gas: low-energy properties and Haldane exclusion statistics,
\href{https://doi.org/10.1103/PhysRevLett.96.210402}{Phys. Rev. Lett. {\bf 96}, 210402 (2006)}.

\bibitem{Mintchev2007}
P. Calabrese and M. Mintchev,
Correlation functions of one-dimensional anyonic fluids,
\href{https://doi.org/10.1103/PhysRevB.75.233104}{Phys. Rev. B {\bf 75}, 233104 (2007)}.

\bibitem{Cabra2007}
R. Santachiara, F. Stauffer, and D. C. Cabra,
Entanglement properties and momentum distributions of hard-core anyons on a ring,
\href{https://doi.org/10.1088/1742-5468/2007/05/L05003}{J. Stat. Mech. {\bf 2007}, L05003 (2007)}.

\bibitem{Pucci2007}
F. M. D. Pellegrino, G. G. N. Angilella, N. H. March, and R. Pucci,
Statistical correlations in an ideal gas of particles obeying fractional exclusion statistics,
\href{https://doi.org/10.1103/PhysRevE.76.061123}{Phys. Rev. E {\bf 76}, 061123 (2007)}.

\bibitem{Averin2007}
O. I. P{\^a}\c tu, V. E. Korepin, and D. V. Averin,
Correlation functions of one-dimensional Lieb-Liniger anyons,
\href{https://doi.org/10.1088/1751-8113/40/50/004}{J. Phys. A: Math. Theor. {\bf 40}, 14963 (2007)}.

\bibitem{Patu2015}
O. I. P{\^a}\c tu,
Correlation functions and momentum distribution of one-dimensional hard-core anyons in optical lattices,
\href{https://doi.org/10.1088/1742-5468/2015/01/P01004}{J. Stat. Mech. {\bf 2015}, P01004 (2015)}.

\bibitem{Freedman2007}
A. Feiguin, S. Trebst, A. W. W. Ludwig, M. Troyer, A. Kitaev, Z. Wang, and M. H. Freedman,
Interacting Anyons in Topological Quantum Liquids: The Golden Chain,
\href{https://doi.org/10.1103/PhysRevLett.98.160409}{Phys. Rev. Lett. {\bf 98}, 160409 (2007)}.

\bibitem{Troyer2008}
S. Trebst, E. Ardonne, A. Feiguin, D. A. Huse, A. W. W. Ludwig, and M. Troyer,
Collective States of Interacting Fibonacci Anyons,
\href{https://doi.org/10.1103/PhysRevLett.101.050401}{Phys. Rev. Lett. {\bf 101}, 050401 (2008)}.

\bibitem{Wang2009}
C. Gils, E. Ardonne, S. Trebst, A. W. W. Ludwig, M. Troyer, and Z. Wang,
Collective States of Interacting Anyons, Edge States, and the Nucleation of Topological Liquids,
\href{https://doi.org/10.1103/PhysRevLett.103.070401}{Phys. Rev. Lett. {\bf 103}, 070401 (2009)}.

\bibitem{Calabrese2008}
R. Santachiara and P. Calabrese,
One-particle density matrix and momentum distribution function of one-dimensional anyon gases,
\href{https://doi.org/10.1088/1742-5468/2008/06/P06005}{J. Stat. Mech. {\bf 2008}, P06005 (2008)}.

\bibitem{Min2008}
Z. Ren-Gui and W. An-Min,
Limiting Case of 1D Delta Anyon Model,
\href{https://doi.org/10.1088/0253-6102/50/6/02}{Commun. Theor. Phys. {\bf 50}, 1265 (2008)}.

\bibitem{Mintchev2009}
B. Bellazzini, P. Calabrese, and M. Mintchev,
Junctions of anyonic Luttinger wires,
\href{https://doi.org/10.1103/PhysRevB.79.085122}{Phys. Rev. B {\bf 79}, 085122 (2009)}.

\bibitem{Roncaglia2011}
T. Keilmann, S. Lanzmich, I. McCulloch, and M. Roncaglia,
Statistically induced phase transitions and anyons in 1D optical lattices,
\href{https://doi.org/10.1038/ncomms1353}{Nat. Commun. {\bf 2}, 361 (2011)}.

\bibitem{Korepin2012}
R. A. Santos, F. N. C. Paraan, and V. E. Korepin,
Quantum phase transition in a multicomponent anyonic Lieb-Liniger model,
\href{https://doi.org/10.1103/PhysRevB.86.045123}{Phys. Rev. B {\bf 86}, 045123 (2012)}.

\bibitem{Santos2015}
S. Greschner and L. Santos,
Anyon Hubbard Model in One-Dimensional Optical Lattices,
\href{https://doi.org/10.1103/PhysRevLett.115.053002}{Phys. Rev. Lett. {\bf 115}, 053002 (2015)}.

\bibitem{Pelster2015}
G. Tang, S. Eggert, and A. Pelster,
Ground-state properties of anyons in a one-dimensional lattice,
\href{https://doi.org/10.1088/1367-2630/17/12/123016}{New J. Phys. {\bf 17}, 123016 (2015)}.

\bibitem{Hao2016}
Y. Hao,
Ground-state properties of hard-core anyons in a harmonic potential,
\href{https://doi.org/10.1103/PhysRevA.93.063627}{Phys. Rev. A {\bf 93}, 063627 (2016)}.

\bibitem{Calabrese2016}
G. Marmorini, M. Pepe, and P. Calabrese,
One-body reduced density matrix of trapped impenetrable anyons in one dimension,
\href{https://doi.org/10.1088/1742-5468/2016/07/073106}{J. Stat. Mech. {\bf 2016}, 073106 (2016)}.

\bibitem{Valencia2016}
J. Arcila-Forero, R. Franco, and J. Silva-Valencia,
Critical points of the anyon-Hubbard model,
\href{https://doi.org/10.1103/PhysRevA.94.013611}{Phys. Rev. A {\bf 94}, 013611 (2016)}.

\bibitem{Valencia2018}
J. Arcila-Forero, R. Franco, and J. Silva-Valencia,
Three-body-interaction effects on the ground state of one-dimensional anyons,
\href{https://doi.org/10.1103/PhysRevA.97.023631}{Phys. Rev. A {\bf 97}, 023631 (2018)}.

\bibitem{Trombettoni2018}
A. Colcelli, G. Mussardo, and A. Trombettoni,
Deviations from off-diagonal long-range order in one-dimensional quantum systems,
\href{https://doi.org/10.1209/0295-5075/122/50006}{Europhys. Lett. {\bf 122}, 50006 (2018)}.

\bibitem{Nayak2008}
C. Nayak, S.H. Simon, A. Stern, M. Freedman, and S. Das Sarma,
Non-Abelian anyons and topological quantum computation,
{\href{http://dx.doi.org/10.1103/RevModPhys.80.1083}{Rev. Mod. Phys. {\bf 80}, 1083 (2008)}}.

\bibitem{Meyer2011}
D. M. Badiane, M. Houzet, and J. S. Meyer,
Nonequilibrium Josephson Effect through Helical Edge States,
\href{https://doi.org/10.1103/PhysRevLett.107.177002}{Phys. Rev. Lett. {\bf 107}, 177002 (2011)}.

\bibitem{Oppen2011}
L. Jiang, D. Pekker, J. Alicea, G. Refael, Y. Oreg, and F. von Oppen,
Unconventional Josephson Signatures of Majorana Bound States,
\href{https://doi.org/10.1103/PhysRevLett.107.236401}{Phys. Rev. Lett. {\bf 107}, 236401 (2011)}.

\bibitem{Furdyna2012}
L. Rokhinson, X. Liu, and J. Furdyna,
The fractional a.c. Josephson effect in a semiconductor-superconductor nanowire as a signature of Majorana particles,
\href{https://doi.org/10.1038/nphys2429}{Nat. Phys. {\bf 8}, 795 (2012)}.

\bibitem{Yacoby2014}
S.-P. Lee, K. Michaeli, J. Alicea, and A. Yacoby,
Revealing Topological Superconductivity in Extended Quantum Spin Hall Josephson Junctions,
\href{https://doi.org/10.1103/PhysRevLett.113.197001}{Phys. Rev. Lett. {\bf 113}, 197001 (2014)}.

\bibitem{Aguado2012}
P. San-Jose, E. Prada, and R. Aguado,
ac Josephson Effect in Finite-Length Nanowire Junctions with Majorana Modes,
\href{https://doi.org/10.1103/PhysRevLett.108.257001}{Phys. Rev. Lett. {\bf 108}, 257001 (2012)}.

\bibitem{MooreRead1991}
G. Moore, N. Read, 
Nonabelions in the fractional quantum Hall effect, 
\href{https://doi.org/10.1016/0550-3213(91)90407-O}{Nucl. Phys. B {\bf 360}, 362 (1991)}.

\bibitem{Volovik1999}
G. E. Volovik, 
Fermion zero modes on vortices in chiral superconductors, 
\href{https://doi.org/10.1134/1.568223}{JETP Lett. {\bf 70}, 609 (1999)}.

\bibitem{Green2000}
N. Read and D. Green,
Paired states of fermions in two dimensions with breaking of parity and time-reversal symmetries and the fractional quantum Hall effect,
\href{https://doi.org/10.1103/PhysRevB.61.10267}{Phys. Rev. B {\bf 61}, 10267 (2000)}.

\bibitem{Liu2004}
K. D. Nelson, Z. Q. Mao, Y. Maeno, Y. Liu,
Odd-Parity Superconductivity in $\text{Sr}_2\text{Ru}\text{O}_4$,
\href{https://doi.org/10.1126/science.1103881}{Science {\bf 306}, 1151 (2004)}.

\bibitem{Rice2004}
M. Rice,
Superfluid Helium-3 Has a Metallic Partner,
\href{https://doi.org/10.1126/science.1106019}{Science {\bf 306}, 1142 (2004)}.

\bibitem{Tewari2006}
S. Das Sarma, C. Nayak, and S. Tewari,
Proposal to stabilize and detect half-quantum vortices in strontium ruthenate thin films: non-Abelian braiding statistics of vortices in a $p_x+ip_y$ superconductor,
\href{https://doi.org/10.1103/PhysRevB.73.220502}{Phys. Rev. B {\bf 73}, 220502 (2006)}.

\bibitem{DasSarma2010}
R. M. Lutchyn, J. D. Sau, and S. Das Sarma,
Majorana fermions and a topological phase transition in semiconductor-superconductor heterostructures,
\href{https://doi.org/10.1103/PhysRevLett.105.077001}{Phys. Rev. Lett. {\bf 105}, 077001 (2010)}.

\bibitem{Sau2010}
J. D. Sau,  R. M. Lutchyn, S. Tewari, and S. Das Sarma,
Generic new platform for topological quantum computation using semiconductor heterostructures,
\href{https://doi.org/10.1103/PhysRevLett.104.040502}{Phys. Rev. Lett. {\bf 104}, 040502 (2010)}.

\bibitem{Alicea2010}
J. Alicea,
Majorana fermions in a tunable semiconductor device,
\href{https://doi.org/10.1103/PhysRevB.81.125318}{Phys. Rev. B {\bf 81}, 125318 (2010)}.

\bibitem{Oppen2010}
Y. Oreg, G. Refael, and F. von Oppen,
Helical liquids and Majorana bound states in quantum wires,
\href{https://doi.org/10.1103/PhysRevLett.105.177002}{Phys. Rev. Lett. {\bf 105}, 177002 (2010)}.

\bibitem{Fisher2011}
J. Alicea, Y. Oreg, G. Refael, F. von Oppen, and M. P. A. Fisher,
Non-Abelian statistics and topological quantum information processing in 1D wire networks,
\href{https://doi.org/10.1038/nphys1915}{Nat. Phys. {\bf 7}, 412 (2011)}.

\bibitem{LossKlinovaja2021}
A.E. Svetogorov, D. Loss, and J. Klinovaja,
Insulating regime of an underdamped current-biased Josephson junction supporting $\mathbb{Z}_3$ and $\mathbb{Z}_4$ parafermions,
\href{https://doi.org/10.1103/PhysRevB.103.L180505}{Phys. Rev. B {\bf 103}, L180505 (2021)}.

\bibitem{Oppen2012}
B. I. Halperin, Y. Oreg, A. Stern, G. Refael, J. Alicea, and F. von Oppen,
Adiabatic manipulations of Majorana fermions in a three-dimensional network of quantum wires,
\href{https://doi.org/10.1103/PhysRevB.85.144501}{Phys. Rev. B {\bf 85}, 144501 (2012)}.

\bibitem{Freedman2006}
M. Freedman, C. Nayak, and K. Walker,
Towards universal topological quantum computation in the $\nu=5/2$ fractional quantum Hall state,
\href{https://doi.org/10.1103/PhysRevB.73.245307}{Phys. Rev. B {\bf 73}, 245307 (2006)}.

\bibitem{Bonderson2010}
P. Bonderson, D. J. Clarke, C. Nayak, and K. Shtengel,
Implementing arbitrary phase gates with Ising anyons,
\href{https://doi.org/10.1103/PhysRevLett.104.180505}{Phys. Rev. Lett. {\bf 104}, 180505 (2010)}.

\bibitem{Lukin2017}
H. Bernien, S. Schwartz, A. Keesling, H. Levine, A. Omran, H. Pichler, S. Choi, A. S. Zibrov, M. Endres, M. Greiner, V. Vuleti{\'c}, and M. D. Lukin,
Probing many-body dynamics on a 51-atom quantum simulator,
\href{https://doi.org/10.1038/nature24622}{Nature {\bf 551}, 579 (2017)}.

\bibitem{Lukin2021}
S. Ebadi, T. T. Wang, H.Levine, A. Keesling, G. Semeghini, A. Omran, D. Bluvstein, R. Samajdar, H. Pichler, W. W. Ho, S. Choi, S. Sachdev, M. Greiner, V. Vuleti{\'c}, and M. D. Lukin,
Quantum phases of matter on a 256-atom programmable quantum simulator,
\href{https://doi.org/10.1038/s41586-021-03582-4}{Nature {\bf 595}, 227 (2021)}.

\bibitem{Gorshkov2020}
F. Liu, S. Whitsitt, P. Bienias, R. Lundgren, A. V. Gorshkov,
Realizing and Probing Baryonic Excitations in Rydberg Atom Arrays,
\href{https://arxiv.org/abs/2007.07258}{arXiv:2007.07258 (2020)}.

\bibitem{Andrews1984}
G. E. Andrews, R.J. Baxter,  and p.J. Forrester, Eight-vertex SOS model and generalized Rogers-Ramanujan-type identities. \href{https://doi.org/10.1007/BF01014383}{J. Stat. Phys. {\bf35}, 193 (1984)}.

\bibitem{Zamolodchikov1985}
A. B. Zamolodchikov and V. A. Fateev, 
Nonlocal (parafermion) currents in two-dimensional conformal field theoryand self-dual critical pointsin $Z_N$-symmetric statistical systems, 
\href{http://www.jetp.ras.ru/cgi-bin/dn/e_062_02_0215.pdf}{Zh. Eksp. Theor. Phys. {\bf 89}, 380 (1985)}.

\bibitem{Fendley2012}
P. Fendley,
Parafermionic edge zero modes in ${\mathbb Z}_{n}$-invariant spin chains,
\href{https://doi.org/10.1088/1742-5468/2012/11/P11020}{J. Stat. Mech. {\bf 2012}, P11020 (2012)}.

\bibitem{Fendley2014}
P. Fendley,
Free parafermions,
\href{https://doi.org/10.1088/1742-5468/2012/11/P11020}{J. Phys. A: Math. Theor. {\bf 47}, 075001 (2014)}.

\bibitem{Fendley2016}
J. Alicea and P. Fendley,
Topological Phases with Parafermions: Theory and Blueprints,
\href{https://doi.org/10.1146/annurev-conmatphys-031115-011336}{Annu. Rev. Condens. Matter Phys. {\bf 7}, 119 (2016)}.

\bibitem{KlinovajaLoss2014}
J. Klinovaja, P. Stano, and D. Loss, 
Exotic states at the edge: Majorana fermions and parafermions, 
\href{http://www.sps.ch/artikel/progresses/exotic-states-at-the-edge-majorana-fermions-and-parafermions-42/.}{ SPG Mitteilungen {\bf 43}, 31 (2014)}.

\bibitem{HutterLoss2016}
A. Hutter and D. Loss,
Quantum computing with parafermions,
\href{https://doi.org/10.1103/PhysRevB.93.125105}{Phys. Rev. B {\bf 93}, 125105 (2016)}.

\bibitem{Cobanera2014}
E. Cobanera and G. Ortiz, 
Fock parafermions and self-dual representations of the braid group, 
\href{https://doi.org/10.1103/PhysRevA.89.012328}{Phys. Rev. A {\bf 89}, 012328 (2014)}. 

\bibitem{Clarke2013}
D. J. Clarke, J. Alicea, and K. Shtengel 
Exotic non-Abelian anyons from conventional fractional quantum Hall states,
\href{https://doi.org/10.1038/ncomms2340}{Nat. Commun. {\bf 4}, 1348 (2013)}.

\bibitem{Hassler2016}
E. Cobanera, J. Ulrich, and F. Hassler, 
Changing anyonic ground degeneracy with engineered gauge fields, 
\href{https://doi.org/10.1103/PhysRevB.94.125434}{Phys. Rev. B {\bf 94}, 125434 (2016)}. 

\bibitem{Mazza2017}
F. Iemini, C. Mora, and L. Mazza, 
Topological phases of parafermions: a model with exactly solvable ground states, 
\href{https://doi.org/10.1103/PhysRevLett.118.170402}{Phys. Rev. Lett. {\bf 118}, 170402 (2017)}. 

\bibitem{Alicea2018}
A. Chew, D. F. Mross, and J. Alicea, 
Fermionized parafermions and symmetry-enriched Majorana modes, 
\href{https://doi.org/10.1103/PhysRevB.98.085143}{Phys. Rev. B {\bf 98}, 085143 (2018)}. 

\bibitem{Cobanera2017}
E. Cobanera, 
Modeling electron fractionalization with unconventional Fock spaces, 
\href{https://doi.org/10.1088/1361-648X/aa718f}{J. Phys.: Condens. Matter {\bf 29}, 305602 (2017)}. 

\bibitem{Lindner2012}
N. H. Lindner, E. Berg, G. Refael, and A. Stern, 
Fractionalizing Majorana fermions: non-Abelian statistics on the edges of Abelian quantum Hall states, 
\href{https://doi.org/10.1103/PhysRevX.2.041002}{Phys. Rev. X {\bf 2}, 041002 (2012)}. 

\bibitem{Burrello2013}
M. Burrello, B. van Heck, and E. Cobanera, 
Topological phases in two-dimensional arrays of parafermionic zero modes, 
\href{https://doi.org/10.1103/PhysRevB.87.195422}{Phys. Rev. B {\bf 87}, 195422 (2013)}.  

\bibitem{Mong2013}
R. S. K. Mong, D. J. Clarke, J. Alicea, N. H. Lindner, P. Fendley, C. Nayak, Y. Oreg, A. Stern, E. Berg, K. Shtengel, and M. P. A. Fisher, 
Universal topological quantum computation from a superconductor-Abelian quantum Hall heterostructure, 
\href{https://doi.org/10.1103/PhysRevX.4.011036}{Phys. Rev. X {\bf 4}, 011036 (2014)}. 

\bibitem{Barkeshli2014}
M. Barkeshli and X.-L. Qi, 
Synthetic topological qubits in conventional bilayer quantum Hall systems, 
\href{https://doi.org/10.1103/PhysRevX.4.041035}{Phys. Rev. X {\bf 4}, 041035 (2014)}. 

\bibitem{ZhangKane2014}
F. Zhang and C.L. Kane, 
Time-reversal-invariant ${\mathbb Z}_{4}$ fractional Josephson effect, 
\href{https://doi.org/10.1103/PhysRevLett.113.036401}{Phys. Rev. Lett. {\bf 113}, 036401 (2014)}. 

\bibitem{KlinovajaLoss20142}
J. Klinovaja and D. Loss,
Time-reversal invariant parafermions in interacting Rashba nanowires
\href{https://doi.org/10.1103/PhysRevB.90.045118}{Phys. Rev. B {\bf 90}, 045118 (2014)}. 

\bibitem{KlinovajaLoss20143}
J. Klinovaja and D. Loss,
Parafermions in an interacting nanowire bundle,
\href{https://doi.org/10.1103/PhysRevLett.112.246403}{Phys. Rev. Lett. {\bf 112}, 246403 (2014)}. 

\bibitem{KlinovajaLoss20144}
J. Klinovaja, A. Yacoby, and D. Loss, 
Kramers pairs of Majorana fermions and parafermions in fractional topological Insulators, 
\href{https://doi.org/10.1103/PhysRevB.90.155447}{Phys. Rev. B {\bf 90}, 155447 (2014)}. 

\bibitem{Schmidt2015}
C. P. Orth, R. P. Tiwari, T. Meng, and T. L. Schmidt, 
Non-Abelian parafermions in time-reversal invariant interacting helical systems, 
\href{https://doi.org/10.1103/PhysRevB.91.081406}{Phys. Rev. B {\bf 91}, 081406 (2015)}. 

\bibitem{Loss2015}
A. Hutter, J. R. Wootton, and D. Loss,
Parafermions in a Kagome lattice of qubits for topological quantum computation,
\href{https://doi.org/10.1103/PhysRevX.5.041040}{Phys. Rev. X {\bf 5}, 041040 (2015)}. 

\bibitem{KlinovajaLoss2015}
J. Klinovaja and D. Loss, 
Fractional charge and spin states in topological insulator constrictions, 
\href{https://doi.org/10.1103/PhysRevB.92.121410}{Phys. Rev. B {\bf 92}, 121410(R) (2015)}. 

\bibitem{Thakurathi2017}
M. Thakurathi, D. Loss, and J. Klinovaja,
Floquet Majorana fermions and parafermions in driven Rashba nanowires,
\href{https://doi.org/10.1103/PhysRevB.95.155407}{Phys. Rev. B {\bf 95}, 155407 (2017)}. 

\bibitem{Laubscher2019}
K. Laubscher, D. Loss, and J. Klinovaja,
Fractional topological superconductivity and parafermion corner states
\href{https://doi.org/10.1103/PhysRevResearch.1.032017}{Phys. Rev. Research {\bf 1}, 032017(R) (2019)}. 

\bibitem{Laubscher2020}
K. Laubscher, D. Loss, J. Klinovaja,
Majorana and parafermion corner states from two coupled sheets of bilayer graphene
\href{https://doi.org/10.1103/PhysRevResearch.2.013330}{Phys. Rev. Research {\bf 2}, 013330 (2020)}. 

\bibitem{Ronetti2021}
F. Ronetti, D. Loss, J. Klinovaja
Clock model and parafermions in Rashba nanowires,
\href{https://doi.org/10.1103/PhysRevB.103.235410}{Phys. Rev. B {\bf 103}, 235410 (2021)}. 

\bibitem{Rossini2019}
D. Rossini, M. Carrega, C. Strinati, and L. Mazza,
Anyonic tight-binding models of parafermions and of fractionalized fermions,
\href{https://doi.org/10.1103/PhysRevB.99.085113}{Phys. Rev. B {\bf 99}, 085113 (2019)}.

\bibitem{Baxter1989}
R. J. Baxter,
A simple solvable ${\mathbb Z}_N$ Hamiltonian,
\href{https://doi.org/10.1016/0375-9601(89)90884-0}{Phys. Lett. A {\bf 140}, 155 (1989).}.

\bibitem{Fradkin1980}
E. Fradkin and L. P. Kadanoff, Disorder variables and para-fermions in two-dimensional statistical mechanics, \href{https://doi.org/10.1016/0550-3213(80)90472-1}{Nucl. Phys. B {\bf 170}, 1 (1980)}.

\bibitem{Trifonov2012}
D.A. Trifonov, Nonlinear fermions and coherent states, \href{https://doi.org/10.1088/1751-8113/45/24/244037}{J. Phys. A: Math. Theor. {\bf 45},  244037 (2012)}.

\bibitem{Mahyaeh2020}
I. Mahyaeh, J. Wouters, and D. Schuricht, Phase diagram of the ${\mathbb Z}_3$-Fock parafermion chain with pair hopping, \href{https://doi.org/10.21468/SciPostPhysCore.3.2.011}{SciPost Phys. Core {\bf 3}, 011 (2020)}.

\bibitem{Ziolkowska2020}
Similar jump operators were studied in e.g. A.A. Ziolkowska, F.H.L. Essler, Yang-Baxter integrable Lindblad equations, \href{doi: 10.21468/SciPostPhys.8.3.044}{SciPost Phys. {\bf 8}, 044, (2020)}.

\bibitem{PTsymmetry}
${\cal PT}$-symmetric Hamiltonians are known to have real spectrum despite being non-Hermitian. For a review, see e.g., C.M. Bender, {\it PT-symmetry in quantum and classical physics}, \href{https://doi.org/10.1142/q0178}{World Scientific, 2019}; R. El-Ganainy, K.G. Makris, M. Khajavikhan, Z.H. Musslimani, S. Rotter, and D.N. Christodoulides, Non-Hermitian physics and PT symmetry, \href{https://doi.org/10.1038/nphys4323}{Nature Phys., {\bf 14}, 11 (2018)}.

\bibitem{Dalibard1992}
J. Dalibard, Y. Castin, and K. M{\o}lmer, Wave-function approach to dissipative processes in quantum optics, \href{https://doi.org/10.1103/PhysRevLett.68.580}{Phys. Rev. Lett. {\bf 68}, 580 (1992)}.

\bibitem{Gardiner1992}
C.W. Gardiner, A.S. Parkins, and P. Zoller, Wave-function quantum stochastic differential equations and quantum-jump simulation methods, \href{https://doi.org/10.1103/PhysRevA.46.4363}{Phys. Rev. A {\bf 46}, 4363 (1992)}.

\bibitem{Plenio1998}
M.B. Plenio and P.L. Knight, The quantum-jump approach to dissipative dynamics in quantum optics, \href{https://doi.org/10.1103/RevModPhys.70.101}{Rev. Mod. Phys. {\bf 70}, 101 (1998)}.

\bibitem{Daley2014}
A.J. Daley, Quantum trajectories and open many-body quantum systems, \href{https://doi.org/10.1080/00018732.2014.933502}{Adv. Phys. {\bf 63}, 77 (2014)}.

\bibitem{Torres2014}
J.M. Torres, Closed-form solution of Lindblad master equations without gain, \href{https://doi.org/10.1103/PhysRevA.89.052133}{Phys. Rev. A {\bf 89}, 052133 (2014)}.

\bibitem{Nakagawa2021}
M. Nakagawa, N. Kawakami, and M. Ueda, Exact Liouvillian spectrum of a one-dimensional dissipative Hubbard model, \href{https://doi.org/10.1103/PhysRevLett.126.110404}{Phys. Rev. Lett. {\bf 126}, 110404 (2021)}.


%
%
%
%
%
%
%
%
%
%
%
%
%
%
%
%
\end{thebibliography}
\end{document}